\documentclass[
  journal=pasa,
  manuscript=BGCE-Student-Paper-Prize, 
  year=2022,
  volume=37,
]{cup-journal}

\usepackage{amsfonts,amsmath,amssymb}

\usepackage{microtype,siunitx,booktabs}
\usepackage{listings}
\usepackage{pgf}
\usepackage{graphicx}
\usepackage{subfigure}
\usepackage{comment}
\usepackage{hyperref}
\usepackage{cleveref}
\usepackage{caption}
\usepackage{xcolor}
\usepackage{multicol}
\usepackage{algorithm}
\usepackage{algpseudocode}

\setcitestyle{numbers,square}
\definecolor{codegreen}{rgb}{0,0.6,0}
\definecolor{codegray}{rgb}{0.5,0.5,0.5}
\definecolor{codepurple}{rgb}{0.58,0,0.82}
\definecolor{backcolour}{rgb}{0.95,0.95,0.95}
\hypersetup{
	unicode=true,          
	pdftoolbar=true,        
	pdfmenubar=true,        
	pdffitwindow=false,     
	pdfstartview={FitH},    
	pdftitle={Tesis Uzmar},    
	pdfauthor={Uzmar Gómez},     
	pdfsubject={Subject},   
	pdfcreator={Uzmar Gómez},   
	pdfproducer={Uzmar Gómez}, 
	pdfkeywords={}, 
	pdfnewwindow=true,      
	colorlinks=true,       
	linkcolor=orange,          
	citecolor=brown,        
	filecolor=brown,      
	urlcolor=brown           
}

\title{GPU Offloading in ExaHyPE Through C++ Standard Algorithms}

\author{Uzmar Gomez}
\affiliation{Department of Computer Science, Durham University, United Kingdom}
\email[Uzmar Gomez]{uzmar.gomez@hotmail.com}

\author{Gonzalo Brito Gadeschi}
\affiliation{NVIDIA, Munich, Germany}

\author{Tobias Weinzierl}
\affiliation{Department of Computer Science, Durham University, United Kingdom}
\vspace{-1cm}

\pgfdeclareimage[height=4.5cm]{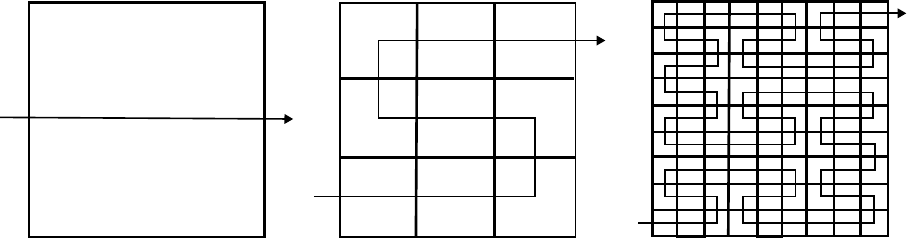}{Figures/cartesiangridrefinement} 
\pgfdeclareimage[width=\textwidth]{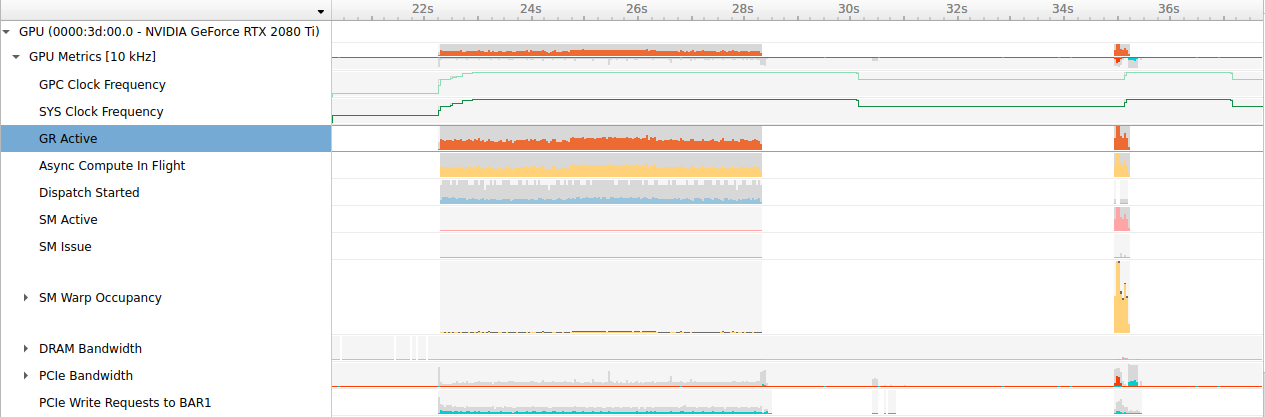}{Figures/nsight_results}

\begin{document}

\begin{abstract}
 The ISO C++17 standard introduces \emph{parallel algorithms}, a parallel programming model promising portability across a wide variety of parallel hardware including multi-core CPUs, GPUs, and FPGAs. 
Since 2019, the NVIDIA HPC SDK compiler suite supports this programming model for multi-core CPUs and GPUs. 
ExaHyPE is a solver engine for hyperbolic partial differential equations for
complex wave phenomena. 
It supports multiple numerical methods including Finite Volumes and
ADER-DG, and employs adaptive mesh refinement with dynamic load balancing via space-filling
curves as well as task-based parallelism and offloading to GPUs.
This study ports ExaHyPE's tasks over blocks of Finite Volumes to the ISO C++
parallel algorithms programming model, and compares its performance and
usability against an OpenMP implementation with offloading via OpenMP target directives.
It shows that ISO C++ is a feasible programming model for non-trivial
applications like our task-based AMR code. 
The realisation is bare of vendor-specific or non-C++ extensions.
It however is slower than its OpenMP
counterpart.
\vspace{-1cm}
\end{abstract}

\section{Introduction}
\label{sec:intro}

%
%
Graphics Processing Units (GPUs) have become the main accelerator for data-parallel processing tasks driving Exascale systems \cite{gpuclusters}. 
Yet, porting applications to GPUs and designing GPU algorithms remains
challenging.
The main HPC programming paradigms for GPU-accelerated systems are ISO standard
languages (C++ and Fortran) that run on GPUs ``automatically'', annotation-based
programming models like OpenMP and OpenACC, and GPU programming languages or
extensions like CUDA or SYCL.
The different approaches have pros and cons.

%
%
The NVIDIA HPC SDK supports automatic offloading of ISO C++ \emph{parallel
algorithms} to multi-core CPUs and GPUs \cite{olsen2019}. Recent publications
demonstrate good performance for mini-applications \cite{lin2022} and NVIDIA's
documentation enlists success stories for larger applications, such as the
multi-physics applications \href{https://www.hpccoe.eu/2021/06/04/m-aia/}{m-AIA} \cite{maia} and \href{https://www.unige.ch/hpfs/research/open-source-software/stlbm}{STLBM} \cite{Latt_2021}.

%
%
The goal of this study is to examine the feasibility of parallelizing a large application to portably target multi-core CPU and multi-GPU systems using plain ISO C++.
ExaHyPE is a solver engine for hyperpolic PDE equations \cite{Reinarz} for complex wave phenomena.
It is a large and complex application with support for adaptive mesh refinement
with dynamic load balancing via space-filling curves.
On top of this MPI+X parallelism for the CPU, it can phrase the updates of 
individual cells as tasks:
We restrict ourselves to the computational kernels of ExaHyPE's Finite Volume
solver for small Cartesian patches in a block-structured AMR context
\cite{Zhang:2022:ISC}---these are embedded into the cells and hance form
tasks---for which pre-existing OpenMP offloading and SyCL versions are
available.
More complex offloading strategies---notably porting execution logic to a GPU---are beyond scope.

The main contributions of this work are discovering how to map individual tasks
to the new ISO C++ \emph{parallel algorithms} programming model, and comparing the productivity, usability, and performance of this programming model against OpenMP.
We also report on the technical challenges we encountered using this novel programming model for offloading ISO C++ to GPUs. 

%
%
This study is organized as follows. Section \ref{section:exahype} describes the problem domain, the numerical scheme, and the ExaHyPE software. Section \ref{section:compute-kernel} discusses the core compute kernel C++ implementation, GPU execution policies, and implementation challenges.
Section \ref{section:results} shows our preliminary results from which we draw
our conclusions.

\section{Block-structured Finite Volumes in ExaHyPE}
\label{section:exahype}


%
%
ExaHyPE is an engine to write simulation codes for hyperbolic, first-order equation systems over functions $\vec{q}\left(\vec{x},t\right)=\left(q_1,\dots,q_s \right)$

\vspace{-0.2cm}
\begin{equation} 
    \frac{\partial \vec{q}}{\partial t} + \sum_{j=1}^d\frac{\partial}{\partial x_j}\vec{f^j}(\vec{q}) = 0,
    \label{equation:exahype:generic-hyperbolic}
\end{equation}
\vspace{-0.4cm}

\noindent
where $\vec{x}\in \mathbb{R}^d$ and the directional terms $\vec{f^j}$ form the flux function. 
In ExaHyPE, users adopt such generic PDE formulations by injecting function
implementations evaluating $f^j(\vec{q})$. How these function evaluations are
used within a numerical scheme, on which core, and when, is hidden from the users. This is the Hollywood principle from \cite{Weinzierl:2019:Peano}.
For the present experiments, we benchmark the Euler equations.
In conservative form, these equations read as

\vspace{-0.2cm}
\begin{equation}
  \frac{\partial}{\partial t}
  \begin{pmatrix}
    \rho\\
    \vec{j}\\
    E_t
  \end{pmatrix}
  +\nabla \cdot 
  \begin{pmatrix}
    \vec{j}  \\
    \frac{1}{\rho}\vec{j}\otimes \vec{j}+pI \\
    \left(E_t+p\right)\frac{1}{\rho}
  \end{pmatrix}
  =\vec{0},
  \label{equation:exahype:Euler}
\end{equation}
\vspace{-0.2cm}

\noindent
where $\vec{j}=\rho\vec{u}$ is the momentum density, $\rho$ is the fluid mass density, $p$ is the pressure, and $E_t$ is the total energy density.

ExaHyPE's spatial discretisation is fixed: The code employs a block-structured
mesh \cite{Dubey:AMR}, resulting from spacetrees \cite{Weinzierl:2019:Peano}: A
cubic computational domain is split into three chunks along each coordinate
axis and this process then is repeated locally and recursively. We end up with a
dynamically adaptive grid of cubes. On top of this spatial discretisation, ExaHyPE offers various numerical schemes. We stick to block-structured Finite Volumes here: Each cube of the grid hosts a small Cartesian patch of $p \times p$ or $p \times p \times p$ volumes. Each patch is supplemented with a halo layer.

It is ExaHyPE's responsibility to maintain, store and traverse the
block-structured mesh.
Per Cartesian patch, the code invokes an update kernel to advance a patch in
time.
This kernel is tailored towards the PDE of interest by the user-provided flux
functions.
After that, it is ExaHyPE's responsibility to keep all (halo) data
consistent.

%
%

\subsubsection*{Parallelisation and GPU offloading}

ExaHyPE employs a traditional non-overlapping domain decomposition realised
through MPI.
Per MPI rank, it then applies a non-overlapping domain decomposition again.
The resulting subdomains are distributed per rank among the threads.

Each thread runs through its subdomain per time step and advances the numerical
solution.
Throughout the traversal, the code prioritises patches along AMR boundaries or
the subdomain boundaries:
These patches are updated immediately while we run throught the domain, as their
output is urgent as it feeds into domain boundary exchange or is
involved in potentially expensive AMR routines.

The remaining patches are held back and put into a thread-local buffer.
Whenever this buffer exceeds a given threshold $N$, we take the $N$ patches and
deploy them as a whole to an accelerator.
All remaining patches are deployed to normal tasks at the end of a traversal to
exploit the host cores.
The classification of patches realises the principle of enclave tasking
\cite{Charrier}, while the buffering uses a concept from \cite{Zhang:2022:ISC}.

\subsubsection*{A generic Finite Volume kernel}

Per volume $\mathcal{C}_i$ within each $p \times p \times p$ patch, 
we use
a Finite Difference approximation of
the time derivative in \eqref{equation:exahype:generic-hyperbolic} subject to a
Finite Volume scheme for the spatial discretisation:

\vspace{-0.2cm}
\begin{equation}
  \frac{Q_i^{n+1} - Q_i^n}{\Delta t} +\frac{F_{i+1/2}^n-F_{i-1/2}^n}{\Delta x} = 0,
  \label{equation:exahype:generic-finite-volume-formulation}
\end{equation}
\vspace{-0.4cm}

\noindent
where $F_{i-1/2}^n$ is the flux over the faces of $\mathcal{C}_i$.
Subsequently, we use Rusanov's approximation to the flux from
(\ref{equation:exahype:generic-finite-volume-formulation}):

\vspace{-0.2cm}
\begin{equation}
  F_{i-1/2}^n = \frac{1}{2}\left[f\left(Q_{i-1}^n\right)+f\left(Q_i^n\right)-a_{i-1/2} \left(Q_i^n-Q_{i-1}^n\right)\right]
  \label{eq:rusanovflux}
\end{equation}
\vspace{-0.4cm}

\noindent
where

\vspace{-0.2cm}
\begin{eqnarray}
  a_{i-1/2} & = & \max\left(\left|f'(q)\right|\right)\ \text{over all }q\text{
  between }Q_{i-1}^n\text{ and }Q_i^n 
  \nonumber
  \\
  & = &
  \max\left(\left|\lambda_k\right|\right).
    \label{equation:exahype:maxeigenval}
\end{eqnarray}
\vspace{-0.4cm}

\noindent
$\lambda_k$ are the eigenvalues of the PDE's Jacobian matrix.

With this generic numerical scheme, we can implement a generic patch update
kernel which is tailored towards (\ref{equation:exahype:Euler}) by
setting the number of quantities of interest to $d+2$ and by injecting two
routines:
one evaluates Euler's flux, the other one returns the maximum eigenvalue.

\section{A C++ GPU computer kernel}
\label{section:compute-kernel}

\begin{algorithm}[htb]
 \caption{
  Compute steps for one patch. A kernel for multiple patches in one rush adds an additional outer loop.
  \label{alg:compsteps}
 }
  \begin{algorithmic}[1]
    \State $Q^{\text{new}} = Q^{\text{old}}$
    \State Compute $a_{i-1/2}Q^{\text{old}}$ where $a_{i-1/2}$ is given by eq. \ref{equation:exahype:maxeigenval}.
    \State $a=\max\left(\lambda_{\max}Q^{\text{old}}\right)$ for face in boundary of $Q^{\text{old}}$.
    \State Update the volume by $a\left(Q^{\text{old}}-Q^{\text{old-1}}\right)$.
    \State Compute the flux given by eq. \cref{eq:rusanovflux}.
    \State Compute the average of the fluxes between $Q^{\text{old}}$ and the adjacent volumes.
    \State Update the volume with the fluxes from their 2d adjacent faces subject to a scaling with $\frac{\Delta t}{\Delta x}$.
    \State Compute the Finite Volume's non-conservative product.
    \State Update the value of $Q^{\text{new}}$.
    \State Compute the patch global maximum eigenvalue.
  \end{algorithmic}
\end{algorithm}

ExaHyPE offers a generic compute kernel implementation based upon functors, but
we also have a templated version handling $N$ patches in one batch.
It consists of a sequence of 
nested loops (Algorithm \ref{alg:compsteps}).
We run over all patches.
Per patch, we execute different steps, whereas each step consists of loops over
cells or faces, respectively, as well as the $s=5$ unknowns of the PDE.


\begin{lstlisting}[
  caption={Allocation of a Cartesian product (iteration space) on the heap.
  This way, it can be used by the GPU.}, captionpos=b,
  escapechar=|,
  basicstyle=\scriptsize\ttfamily,
  label={allocatingcode}
  ]
// Detecting the cartesian product type automatically
using cart_prod_3d = decltype(tl::views::cartesian_product(
 std::views::iota(0,1), std::views::iota(0,1),
 std::views::iota(0,1)
));

// Allocating in heap
auto* range2d_cell = new cart_prod_3d;

// Defining components
*range2d_cell = tl::views::cartesian_product(
 std::views::iota(0,numberOfVolumesPerAxisInPatch), 
 std::views::iota(0,numberOfVolumesPerAxisInPatch),
 std::views::iota(0,hostPatchData.numberOfCells)
);

...

delete[] range2d_cell;
\end{lstlisting}

\subsubsection*{SIMD loop traversals and functors}

\begin{lstlisting}[
  caption={Standard \lstinline{std::for_each} algorithm offloaded to GPU.},
  captionpos=b,
  escapechar=|,
  basicstyle=\scriptsize\ttfamily,
  label={foreachcode}
  ]
std::for_each(|\label{line:foreach}|
  std::execution::par_unseq,|\label{line:expol}|
  range2d_cell->begin(),|\label{line:beginit}| range2d_cell->end(),|\label{line:endit}|
[
  =,|\label{line:beginalstructheap}|
  QIn=hostPatchData.QIn,
  cellCentre=hostPatchData.cellCentre,
  cellSize=hostPatchData.cellSize,
  t=hostPatchData.t,   dt=hostPatchData.dt,
  QOut=hostPatchData.QOut|\label{line:endalstructheap}|
](auto ids){
  auto [x,y,patchIndex] = ids; |\label{line:sepidx}|

  // function call has to refer to a template, i.e. 
  // implementation has to be in header. Otherwise,  
  // compiler does not know that it has to create a 
  // GPU variant of function.
    
  // step (1) from algorithm blueprint
  internal::copySolution_LoopBody<Solver>(
    QIn[patchIndex],         QInEnumerator,
    cellCentre[patchIndex],  cellSize[patchIndex],
    patchIndex,              volumeIndex2d(x,y),
    t[patchIndex],           dt[patchIndex],
    QOut[patchIndex],        QOutEnumerator
  );

  [...]    
  // steps (2--10) from algorithm blueprint
}|\label{line:endlambdafunc}|
);
\end{lstlisting}

The original kernel implementation used virtual-function dynamic dispatch and separately compiled implementations - i.e. in separate .cpp files - to reduce compilation times and improve developer productivity.
However, the NVIDIA implementation of ISO C++ does not support virtual-function dispatch in device code and separate compilation.
We hence replace virtual-function dispatch in ExaHyPE with template functions
defined in header files and static dispatch (Code~\ref{foreachcode}).
In the case of the maximum in (Algorithm~\ref{equation:exahype:maxeigenval}), we
replace the standard C++ function invocation by a manual comparison.

While the NVIDIA implementation supports separate compilation via non-standard CUDA-like \lstinline{__host__ __device__} annotations, these require the programmer to choose between using a non-standard extension or moving implementation details to header files, which significantly increases compile-times.
A manual implementation of maximum is not ``genuine C++''.

\subsubsection*{Multi-dimensional iteration spaces with Cartesian Product}

Efficient GPU kernels require large iteration spaces of collapsed loops to harvest all 
parallelism available in modern accelerators.
Hence,
a single C++ parallel algorithm that collapses several \lstinline{for} loops is built 
in our Code \ref{allocatingcode}.
It combines multiple
\href{https://en.cppreference.com/w/cpp/ranges/iota_view}{\lstinline{std::views::iota}}
structures using C++23's \href{https://en.cppreference.com/w/cpp/ranges/cartesian_product_view}{\lstinline{std::views::cartesian_product}}.
This implementation takes pointers to the host-thread stack. 
However, according to the HPC SDK documentation \cite{hpcsdkdoc}, only the host heap can be accessed from GPU kernels on the hardware platform used in this study.
As a consequence, our code \ref{allocatingcode} had to allocate the underlying Cartesian product on the
heap.

The absence of an implementation of Cartesian product in the C++ standard
library used in this study requires us to include a preliminary implementation that is not tuned for accelerators.
The heap placement is solely due to technical constraints.

\subsubsection*{Data transfer}

The ISO C++ programming model relies on unified memory to enable the accelerator devices to access all program memory.
The NVIDIA implementation is non-conforming on non-coherent hardware platforms like the one used in this study: devices are only allowed to access host heap memory.
While ExaHyPE's data is mainly stored on the heap, pointers to this data are conveniently aggregated in structs like \lstinline{hostPatchData}, which are often stored on the host's stack.
Device code is not allowed to access these values to load pointers to heap
data.
ExaHyPE hence captures these pointers explicitly in the lambdas passed to the
parallel algorithms.
Lambda captures are copied to the GPU on kernel launch, while the heap data these pointers refer to is migrated on-demand as the GPU threads access it. 
This data stays on the GPU until the CPU access it again, in which case it is automatically migrated back.
There is no portable C++ API for prefetching heap data between the host and the device.

\subsubsection*{Software stack}

Our work uses the NVIDIA HPC SDK 22.3 which leverages the system's C++ standard library (\lstinline{libstdc++}) but uses its own \emph{C++ parallel algorithms} implementation based on the Thrust library.
Many HPC systems do not default to a sufficiently new version of libstdc++, which requires us to manually pick a different one via compiler flags. 
We reported some genuine \lstinline{nvc++} bugs - e.g. ~\cite{bugnvidia} - which will be solved in future releases.
\section{Results}
\label{section:results}


\begin{figure*}[htb]
    \centering
    \includegraphics[width=\textwidth]{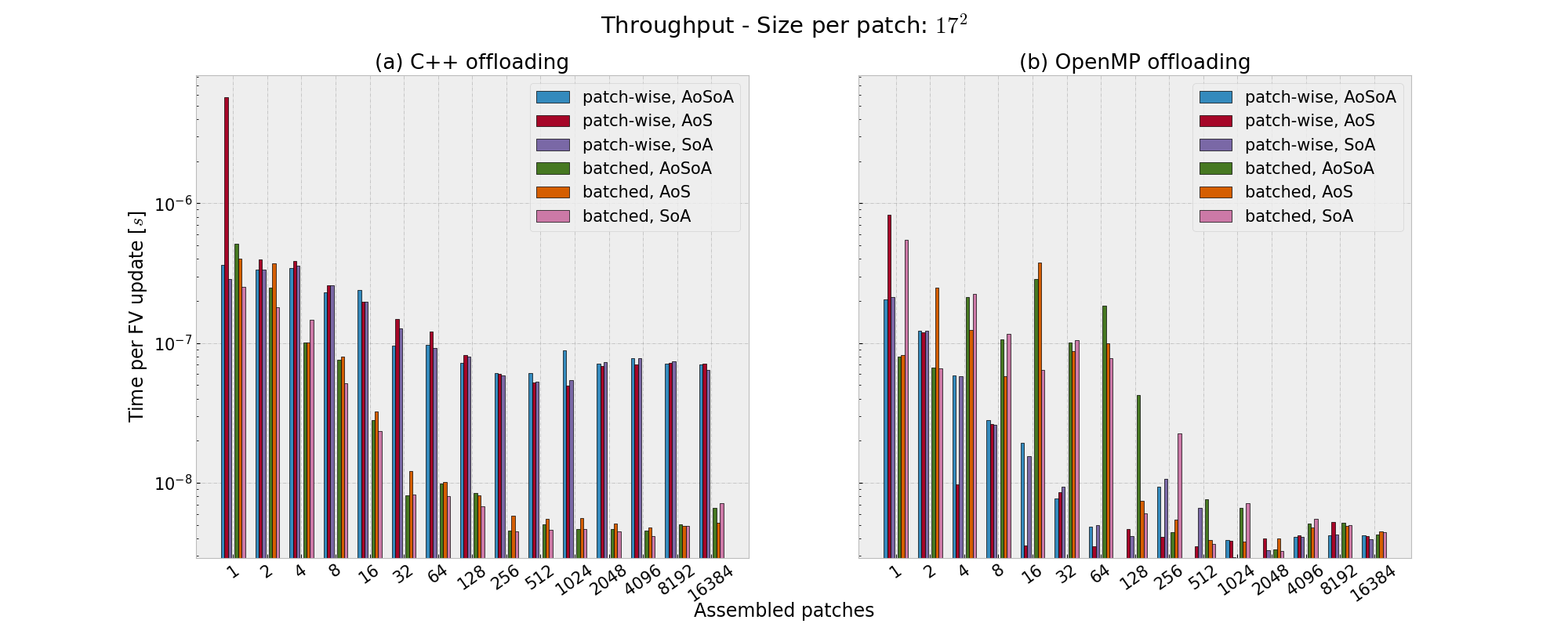}
    \caption{
      Time per Finite Volume update for different implementation
      variants and choices of the number of patches $N$ handled 
      per kernel invocation.
      \label{fig:throughputresults}
      \vspace{-0.4cm}
    }
\end{figure*}


Fig. \ref{fig:throughputresults} shows the benchmark results of our compute kernels for $p=17$ in a 2d setup on a NVIDIA 2080TI for different sizes of patch sets $N$.
All timings include all data transfers between host and device.
The kernels over $N$ patches are available as patch-wise implementation where
the outermost Cartesian loop runs over the patches, and a batched version, where
we run through the individual steps of Algorithm \ref{alg:compsteps} and invoke
a loop over all patches,  all volumes, and all unknowns per step.

Both kernel variants exploit the symmetries from (\ref{eq:rusanovflux}):
The face flux enters the left and right adjacent volumes' solutions.
We compute $f$ (as well as the eigenvalues) per volume and determine
the linear combination of these values from (\ref{eq:rusanovflux}) redundantly
for the left and right adjacent cell.
The intermediate $f$ and $\lambda$ values can be held as Array of
Structures (AoS), Structure of Arrays (SoA) or Array of Structures of Arrays
(AoSoA).

The batched vs.~patch-wise flavours result from a permutation of the
loops within the Cartesian product (Code \ref{allocatingcode}).
However, our AMR code employs local and adaptive timestepping and hence requires
the maximum eigenvalue per patch as additional output per kernel invocation. 
We need a per-patch reduction.
This reduction is performed by using a sequential \lstinline{std::for_each} loop
to avoid data races, but it could be parallelized by using another standard algorithm or atomic memory operations.

%
%
Our results show that the batched variant
is always faster than the patch-wise variant for the C++ implementations. 
It is advantageous to split an algorithm into its steps and to use a large
iteration space per step.
As the time per Finite Volume update decreases with the number of
patches offloaded, we may assume that the GPU kernel does a good job in keeping
the data in-between two steps on the GPU.
Data data layout of temporary data within the computation has limited
performance impact.

%
%
For OpenMP, we see that it can be advantageous to run over the patches
one-by-one.
We assume that different patches in this case end up on different SMs on the
GPU, and run more or less independent of each other.
Overall, the OpenMP port is significantly faster---notably for small
$N$-counts---than the C++ port.
NSight studies suggest that this is due to a bespoke OpenMP memory management,
where we have explicit control via device pointers when data are offloaded. 
The C++ version in contrast has the on-demand memory transfers baked into the implementation.

%
%
Though our software is ready to allow multiple tasks to offload to the GPU
simultaneously, neither C++ nor OpenMP did support this feature.
Both ran into segmentation faults.
\section{Outlook and conclusion}
\label{section:conclusion}

GPU offloading via C++ execution policies within C++ standard algorithms is, in
principle, feasible for non-trivial test cases as our task-based AMR code.
The resulting realisation is bare of specific (OpenMP) extensions and clean;
though this is a subjective assessment.
Performance-wise, the C++ port is slower than its OpenMP counterpart.
It is hard to pinpoint a reason for this behaviour.
The two-dimensional Euler equations have a very low arithmetic intensity.
The performance discrepancies might disappear for more complex calculations such as the ADER-DG ExaHyPE solvers; a phenomenon to be studied in the future.

It is early days to comment on the pros and cons of C++ GPU offloading. 
Besides obvious teething problems, our work highlights particular
challenge for developers:
While C++ may be syntactically and semantically correct, its GPU transfer
might fail (as data is not allocated ``artifically'' on the heap, e.g.), or
conform by running the code sequentially, violating the intention of the programmer to offload computations to an accelerator.
On top of this, the C++ community needs a portable alternative to NVIDIA's vendor-specific
\lstinline{__host__ __device__} annotations for separate compilation.
In the ideal case, multi-pass compilation or link-time compilation for GPUs
would make these annotations unnecessary.

As alternative to this automatisation, 
the C++ language might supplement existing execution
policies with declarative GPU offloading policies.
Using such policies would render otherwise correct loop bodies invalid if they
do not fit to the GPU.
It is not clear how this is properly reflected within the language, or
exclusively should be handled by the compiler throughout translation.

\bibliography{bibliography}

\begin{thebibliography}{10}

\bibitem{gpuclusters}
Volodymyr~V. Kindratenko, Jeremy~J. Enos, Guochun Shi, Michael~T. Showerman,
  Galen~W. Arnold, John~E. Stone, James~C. Phillips, and Wen-mei Hwu.
\newblock Gpu clusters for high-performance computing.
\newblock In {\em 2009 IEEE International Conference on Cluster Computing and
  Workshops}, pages 1--8, 2009.

\bibitem{olsen2019}
David Olsen.
\newblock C++17 parallel algorithms on nvidia gpus with pgi c++.
\newblock In {\em GPU Technology Conference}, 2019.

\bibitem{lin2022}
Wei-Chen Lin, Tom Deakin, and Simon McIntosh-Smith.
\newblock Evaluating iso c++ parallel algorithms on heterogeneous hpc systems.
\newblock In {\em 13th IEEE International Workshop on Performance Modeling,
  Benchmarking and Simulation of High Performance Computer Systems}, 2022.

\bibitem{maia}
European Exascale~Computing Center~of Excellence.
\newblock {Research on AI- and Simulation-Based Engineering at Exascale}.
\newblock \url{https://www.hpccoe.eu/2021/06/04/m-aia/}, 2021.
\newblock Last updated: Sep 1, 2022.

\bibitem{Latt_2021}
Jonas Latt, Christophe Coreixas, and Joël Beny.
\newblock Cross-platform programming model for many-core lattice boltzmann
  simulations.
\newblock {\em {PLOS} {ONE}}, 16(4):e0250306, apr 2021.

\bibitem{Reinarz}
Anne Reinarz, Dominic~E. Charrier, Michael Bader, Luke Bovard, Michael Dumbser,
  Kenneth Duru, Francesco Fambri, Alice-Agnes Gabriel, Jean-Matthieu Gallard,
  Sven Köppel, Lukas Krenz, Leonhard Rannabauer, Luciano Rezzolla, Philipp
  Samfass, Maurizio Tavelli, and Tobias Weinzierl.
\newblock Exahype: An engine for parallel dynamically adaptive simulations of
  wave problems.
\newblock {\em Computer Physics Communications}, 254:107251, 2020.

\bibitem{Zhang:2022:ISC}
Han Zhang, Tobias Weinzierl, Holger Schulz, and Baojiu Li.
\newblock {Spherical accretion of collisional gas in modified gravity I:
  self-similar solutions and a new cosmological hydrodynamical code}.
\newblock {\em Monthly Notices of the Royal Astronomical Society},
  515(2):2464--2482, 07 2022.

\bibitem{Weinzierl:2019:Peano}
T.~Weinzierl.
\newblock The peano software---parallel, automaton-based, dynamically adaptive
  grid traversals.
\newblock {\em ACM Transactions on Mathematical Software}, 45(2):14, 2019.

\bibitem{Dubey:AMR}
A.~Dubey, A.~S. Almgren, J.~B. Bell, M.~Berzins, S.~R. Brandt, G.~Bryan,
  P.~Colella, D.~T. Graves, M.~Lijewski, F.~L{\"{o}}ffler, B.~O'Shea,
  E.~Schnetter, B.~van Straalen, and K.~Weide.
\newblock A survey of high level frameworks in block-structured adaptive mesh
  refinement packages.
\newblock {\em Journal of Parallel and Distributed Computing},
  74(12):3217--3227, 2016.

\bibitem{Charrier}
Dominic~Etienne Charrier, Benjamin Hazelwood, and Tobias Weinzierl.
\newblock Enclave tasking for dg methods on dynamically adaptive meshes.
\newblock {\em SIAM Journal on Scientific Computing}, 42(3):C69--C96, 2020.

\bibitem{hpcsdkdoc}
{NVIDIA HPC Compilers. C++ Parallel Algorithms}.
\newblock
  \url{https://docs.nvidia.com/hpc-sdk/compilers/c++-parallel-algorithms/index.html},
  2022.

\bibitem{bugnvidia}
{NVIDIA Bug report}.
\newblock \url{https://nvbugs/3706250}, 2022.

\end{thebibliography}

\appendix

\newpage
\section{OpenMP reference implementation}

\begin{lstlisting}[
  float=*t,
  caption={Allocation and deallocation of variables on the device for OpenMP.},
  captionpos=b,
  escapechar=|,
  basicstyle=\scriptsize\ttfamily,
  label={allocatingcodeopenmp}
  ]
const int numberOfCells            = hostPatchData.numberOfCells;
double*   rawPointerToCellCentres  = hostPatchData.cellCentre[0].data();
double*   rawPointerToCellSizes    = hostPatchData.cellSize[0].data();
double*   t                        = hostPatchData.t;
double*   dt                       = hostPatchData.dt;
double*   maxEigenvalue            = hostPatchData.maxEigenvalue;

#pragma omp target enter data map(to:rawPointerToCellCentres[0:hostPatchData.numberOfCells*Dimensions]) device(targetDevice) |\label{beginallocation}|
#pragma omp target enter data map(to:rawPointerToCellSizes[0:hostPatchData.numberOfCells*Dimensions]) device(targetDevice)
#pragma omp target enter data map(to:mappedPointersToQIn[0:hostPatchData.numberOfCells]) device(targetDevice)
#pragma omp target enter data map(to:mappedPointersToQOut[0:hostPatchData.numberOfCells]) device(targetDevice)
#pragma omp target enter data map(to:t[0:hostPatchData.numberOfCells]) device(targetDevice)
#pragma omp target enter data map(to:dt[0:hostPatchData.numberOfCells]) device(targetDevice)
#pragma omp target enter data map(alloc:maxEigenvalue[0:hostPatchData.numberOfCells]) device(targetDevice) |\label{endallocation}|

[...]

#pragma omp target exit data map(delete:rawPointerToCellCentres[0:hostPatchData.numberOfCells*Dimensions]) device(targetDevice) |\label{begindeallocation}|
#pragma omp target exit data map(delete:rawPointerToCellSizes[0:hostPatchData.numberOfCells*Dimensions]) device(targetDevice)
#pragma omp target exit data map(delete:mappedPointersToQIn[0:hostPatchData.numberOfCells]) device(targetDevice)
#pragma omp target exit data map(delete:mappedPointersToQOut[0:hostPatchData.numberOfCells]) device(targetDevice)
#pragma omp target exit data map(delete:t[0:hostPatchData.numberOfCells]) device(targetDevice)
#pragma omp target exit data map(delete:dt[0:hostPatchData.numberOfCells]) device(targetDevice)
#pragma omp target exit data map(from:maxEigenvalue[0:hostPatchData.numberOfCells]) device(targetDevice) |\label{enddeallocation}|
\end{lstlisting}

Our OpenMP vanilla version hosts the data of $N$ (number of cells)
patches within a struct on the host.
Data here comprises the actual Finite Volume data as well as meta data such as
cell (patch) sizes and positions.

The OpenMP variant maps these data explicitly onto the
GPU prior to the kernel invocation (Code \ref{allocatingcodeopenmp}).
An analogous loop over the individual patches transfers the Finite Volume data
per patch (now shown).

\begin{lstlisting}[
  caption={Main loop body for the OpenMP code. Equivalent to Code \ref{foreachcode}.},
  captionpos=b,
  escapechar=|,
  basicstyle=\scriptsize\ttfamily,
  label={forloopopenmp}
  ]
#pragma omp target teams distribute parallel for simd collapse(3) device(targetDevice)
for (int patchIndex=0; patchIndex<numberOfCells; patchIndex++)
for (int y = 0; y < numberOfVolumesPerAxisInPatch; y++)
for (int x = 0; x < numberOfVolumesPerAxisInPatch; x++) {
  
    GPUCellData patchData; |\label{begincreategpucelldata}|
    patchData.numberOfCells     = numberOfCells;
    patchData.cellCentre        = (tarch::la::Vector<Dimensions,double>*)(rawPointerToCellCentres);
    patchData.cellSize          = (tarch::la::Vector<Dimensions,double>*)(rawPointerToCellSizes);
    patchData.QIn               = mappedPointersToQIn;
    patchData.QOut              = mappedPointersToQOut;
    patchData.t                 = t;
    patchData.dt                = dt;
    patchData.maxEigenvalue     = maxEigenvalue; |\label{endcreategpucelldata}|

    [...]    
    
    copySolution_LoopBody<Solver>(
        patchData.QIn[patchIndex],
        QInEnumerator,
        patchData.cellCentre[patchIndex],
        patchData.cellSize[patchIndex],
        patchIndex,
        volumeIndex2d(x,y),
        patchData.t[patchIndex],
        patchData.dt[patchIndex],
        patchData.QOut[patchIndex],
        QOutEnumerator
    );
    
    [...]
}
\end{lstlisting}

Our OpenMP reference implementation realises the nested Cartesian iteration
space via pure nested for loops (Code \ref{forloopopenmp}).
They are collapsed via a pragma.
In this sense, the realisation is closer to C than its C++ GPU offloading
counterpart.

Within the loop body, the code reconstructs the host struct manually using
device pointers, i.e.~it creates a struct counterpart on the GPU and befills its
translated GPU pointers manually.
Alternatively, a realisation could have supplemented the struct hosting the $N$
patch data with an OpenMP mapping.

The function \lstinline{copySolutionAndAddSourceTerm_LoopBody} and other compute
step realisations from Algorithm \ref{alg:compsteps} are templates and therefore
found in headers.
Consequently, the compiler can directly inline them and deduce from the context
that they have to be translated for the GPU.

\section{Test environment}

We provide a docker image to reproduce the results obtained in this paper.  
The custom image is based on the latest HPC SDK compiler, and it includes some
Python libraries and other necessary software to run ExaHyPE and its
underlying Peano AMR framework.
If the software is going to be run on a GPU, use the following command to pull the image and start a bash terminal

\begin{lstlisting}[
  language={},
  caption={},
  captionpos=b,
  escapechar=|,
  basicstyle=\scriptsize\ttfamily,
  ]
docker run --gpus all -it ghcr.io/uzmargomez/peano:hpcsdk22.9 bash
\end{lstlisting}

\noindent
Without access to a GPU, the \lstinline{--gpus all} flag has to be omitted.
In the container's terminal, we next run the commands

\begin{lstlisting}[
  language={},
  captionpos=b,
  escapechar=|,
  basicstyle=\scriptsize\ttfamily,
  ]
cd /

git clone https://gitlab.lrz.de/hpcsoftware/Peano.git && cd Peano && git checkout gpus && git checkout c57ef3b5

export PYTHONPATH=/Peano/python
  
libtoolize && aclocal && autoconf && autoheader && cp src/config.h.in . && automake --add-missing

./configure --enable-exahype --enable-blockstructured --enable-loadbalancing --with-gpu=cpp --with-multithreading=cpp CXX=$NVCPP CC=$NVCPP CXXFLAGS="-O4 -std=c++20 -stdpar=gpu -gpu=cc75" LDFLAGS="-stdpar=gpu -gpu=cc75"

make -j4

cd benchmarks/exahype2/euler/fv-kernel-benchmarks 

python3 euler-fv-rusanov-kernel-benchmarks.py -a 1e-4

make -j8

./fv-kernel-benchmarks
\end{lstlisting}

\noindent
If a GPU is not available, the HPC SDK compiler can use a multicore CPU to parallelize standard algorithms. 
In this case, the image's application benchmarks solely the CPU kernel
performance.
In order to do this, we change the flags \lstinline{-stdpar=gpu} to
\lstinline{-stdpar=multicore}.

\end{document}